\documentclass[12pt]{article}

\advance\textwidth by 0.5in
\advance\hoffset by -0.25in

\newcommand{\be}{\begin{eqnarray}}
\newcommand{\ee}{\end{eqnarray}}
\newcommand{\non}{\nonumber\\}
\newcommand{\GeV}{\hbox{GeV}}

\usepackage{epsf}
\begin{document}
\title {\bf Polarized Gluon Distribution Function from $\eta'$ Production}

\author
{
 Sangyong Jeon \\
 {\small\it Department of Physics, McGill University, 
   Montreal, QC H3A-2T8, Canada}\\
{\small\rm and}\\
 {\small\it RIKEN-BNL Research Center, BNL, Upton, NY 11973, USA} 
\\
\\
Jamal Jalilian-Marian \\
 {\small\it Physics Department, BNL, Upton, NY 11973, USA}\\
}

 \maketitle

 \begin{abstract}
Using the recently proposed $gg\eta'$ effective vertex, 
we investigate the production of $\eta'$ from gluon fusion in polarized 
$pp$ collisions. We show that by measuring $A_{LL}$ in $\eta'$ production, 
one can extract the polarized gluon distribution 
$\Delta G(x,Q^2)$ at  $Q^2 \sim 1\,\GeV^2$ and in a wide range of $x$.
\end{abstract}

 \newpage

 \section{Introduction}

The Relativistic Heavy Ion Collider (RHIC) at Brookhaven National 
Laboratory will collide polarized protons at a center 
of mass energy of $\sqrt{s}=250\,\GeV$ in very near future 
(for a review of the spin
program at RHIC, see \cite{bssv}). One of the most important
goals of the experiment is to measure the poorly known polarized 
gluon distribution in a proton. In general, polarized distributions 
will be measured at RHIC in a more extended kinematic region and with 
higher accuracy than the previous experiments. 

Even though Deep Inelastic Scattering of electrons on
protons is the best process in which to measure QCD structure 
functions and the parton distribution functions, it is also possible 
to extract the parton distribution functions from Drell-Yan, 
direct photons, $J/\psi$'s, etc.
Recently, we proposed \cite{js1} measuring $\eta'$ in (unpolarized) 
proton-nucleus collisions as a way to extract the gluon distribution 
function in nuclei. This process allows one to directly 
(without any deconvolution) extract the nuclear gluon distribution 
function.  Furthermore, the accessible kinematic region in $x$
in this measurement is orders of magnitude smaller than 
other processes in hadronic collisions. One can also investigate
the possibility of the restoration of $U_A(1)$ symmetry at zero temperature
due to high gluon density effects in nuclei \cite{jks}. Production
of $\eta'$ in heavy ion collisions was considered in \cite{sj}.

In this short note, we extend our previous work on $\eta'$ production
in proton-nucleus collisions to {\em polarized} proton-proton collisions 
at RHIC and estimate the double spin asymmetry $A_{LL}$\footnote{
The possibility of using $\eta'$ at low $P_t$ to probe the
polarization of gluons inside the proton was first mentioned in 
Ref.\protect\cite{Castoldi:1997fn}.}.

We show that
the predicted asymmetry is large enough to be experimentally 
measurable in a wide range of rapidity. We predict the asymmetry
$A_{LL}$ to be $\sim 0.001$ at mid rapidity using the available
parameterization of the polarized gluon distribution functions.
Alternatively, by measuring the asymmetry $A_{LL}$ in the mid
rapidity region, one can probe the polarized gluon distribution
function at $x \sim 0.004$ and $Q^2= 1\,\GeV^2$.

\section{$\eta'$ production cross section}

The $\eta'$ meson is much heavier than its fellow pseudo-scalar mesons,
$\pi, K$ and $\eta$.
This fact has intrigued the physics community for decades. It is
now understood that $\eta'$ gets most of its mass through
instantons and breaking of $U_A(1)$ symmetry due to the triangle
anomaly \cite{th,wit}. Atwood and Soni used this
anomaly to propose the following effective vertex for gluon-gluon-$\eta'$
coupling in \cite{as}  
 \be
 T^{\alpha\beta}_{ab}(p,q,P)
 = H(p^2, q^2, P^2)\, \delta_{ab} \,
 \epsilon_{\mu\nu\lambda\gamma}\, 
 p^\mu \, q^\nu \,
 \epsilon^{\lambda}(p,\alpha) \,
 \epsilon^{\gamma}(q,\beta)
 \label{eq:vertex}
 \ee
$p$ and $q$ are the momanta of the gluons while $P$ is the momentum 
of the produced $\eta'$.
Here, $\epsilon_{\lambda}(p,\alpha)$ is the
polarization vector of a gluon with momentum $p$ and helicity 
$\alpha=\pm 1$.  The Kronecker delta $\delta_{ab}$ indicates that 
only the color singlet combination of the gluons contribute.
In Ref.\cite{as}, the form factor $H(p^2, q^2, P^2)$ was estimated to be
$1.8 \,\GeV^{-1}$ in the limit where
the incoming gluons and the produced $\eta'$ are on-shell.
Using this effective vertex, it is easy to calculate the differential 
cross section for $gg\rightarrow \eta'$ which is given by
\cite{js1}
\be
d\hat{\sigma}^{gg\rightarrow \eta'}_{ab,\alpha\beta}
= {1 \over 4 \sqrt{(p\cdot q)^2}} 
|T_{ab}^{\alpha\beta}|^2 (2\pi)^4 \delta^4(P-p-q) 
{d^3 P \over (2\pi)^3 2 E_P}
\label{eq:dsig}
\ee 
with $p^2 = q^2 = 0$ and $P^2 = M_{\eta'}^2$.

We will now consider the collision of longitudinally polarized protons 
at RHIC. The double spin asymmetry $A_{LL}$ is defined as (for an 
introduction to spin effects in high energy collisions, see 
\cite{el,chjr,ael})
\be
A_{LL}\equiv {d\sigma_{++} - d\sigma_{+-} \over d\sigma_{++} + d\sigma_{+-}}
\equiv {d\Delta\sigma \over 2 d\sigma}
\label{eq:Adef}
\ee
where $ d\sigma_{++}$ denotes the cross section where both protons have their
spins parallel to their momenta while  $ d\sigma_{+-}$ denotes the
cross section when one proton has its spin anti-parallel to its momentum.
The unpolarized proton-proton differential cross section is denoted by
$d\sigma$.

The differential cross section $ d\Delta\sigma $ is related to the
polarized gluon distribution function through 
\be
d\Delta\sigma^{pp\rightarrow \eta'X} = \int dx_1 dx_2 \Delta G(x_1,Q_f^2)  
\Delta G(x_2,Q_f^2)  d\Delta\hat{\sigma}^{gg\rightarrow \eta'}
\label{eq:colin}
\ee
where $x_1$ and $x_2$ are the momentum fractions of the incoming
gluons, $Q^2_f$ is the factorization scale and 
$ d\Delta\hat{\sigma} \equiv d\hat{\sigma}_{++} -  d\hat{\sigma}_{+-}$.
Here $d\hat{\sigma}_{++}$ is the differential cross section for scattering
of two positive helicity gluons averaged over the color. Explicitly, 
\be
d\hat{\sigma}_{++} = 
{1\over (N_c^2 - 1)^2}\sum_{a,b} d\hat{\sigma}^{gg\rightarrow \eta'}_{ab,++}
\label{eq:dsig++}
\ee
and similarly for $d\hat{\sigma}_{+-}$. 
However, since $\eta'$ is a pseudo-scalar,
it is clear that only positive helicity gluons will contribute. 
In other words,
\be
d\Delta \sigma^{gg\to\eta'} = d\hat{\sigma}_{++}
\ee
Calculation of $ |T_{ab}^{++}|^2$
is straightforward and gives
\be
|T_{ab}^{++}|^2 = \delta_{ab}{M_{\eta'}^4 H_0^2\over 4}
\ee
where we defined $H_0 = H(0,0,M_{\eta'}^2)$.
The
differential cross section for inclusive $\eta'$ production in
polarized proton-proton collisions is then%
\footnote{%
 The unpolarized cross-section calculated in Ref.\cite{js1}
 is
 \be
 {d\sigma^{pp\rightarrow \eta' X} \over dx_L} 
 & = &
 {\pi\, H_0^2 \over 64\, x_E}\,
 x_+ G(x_+,Q^2_f)\, x_- G(x_-,Q^2_f)
 \ee
 which is the average of the polarized cross-sections
 $d\sigma_{++}^{pp\rightarrow \eta' X}/dx_L$
 and
 $d\sigma_{+-}^{pp\rightarrow \eta' X}/dx_L$.
 }

\be
{d\Delta\sigma^{pp\rightarrow \eta' X} \over dx_L} =
{\pi\, H_0^2 \over 32\, x_E}\,
x_+ \Delta G(x_+,Q^2_f)\, x_- \Delta G(x_-,Q^2_f)
\label{eq:diffcs}
\ee
where $x_L = 2P_z/\sqrt{s}$ and $x_{E} = 2E_P/\sqrt{s}$ are the 
momentum and energy fractions of the produced $\eta'$ respectively
and 
\be
x_\pm \equiv {x_E \pm x_L\over 2} = {E_P \pm P_z\over \sqrt{s}}
\label{eq:xpmdef}
\ee 

Using (\ref{eq:diffcs}) in (\ref{eq:Adef}) leads to the following result 
for the asymmetry $A_{LL}$ 
\be
A_{LL}= {\Delta G(x_+,Q^2_f)\, \Delta G(x_-,Q^2_f)
\over  G(x_+,Q^2_f)\, G(x_-,Q^2_f) }
\label{eq:Aresult}
\ee 
Equation (\ref{eq:Aresult}) is our main result. It simply relates
the measured double spin asymmetry to the polarized gluon distribution
in a proton. 

\section{Results}

To estimate the cross-sections and the theoretical uncertainties,
we use following two sets of gluon distribution functions.
The first set consists of 
the MRST99\cite{Martin:1999ww} unpolarized gluon distribution function
and LSS2001 \cite{Leader:2001kh} polarized gluon distribution function 
which is based on MRST99.
The second set consists of GRV98\cite{Gluck:1998xa}
unpolarized gluon
distribution function and GRSV2000\cite{Gluck:2000dy} polarized gluon
distribution function which is based on GRV98. 
The minimum $Q^2$ allowed by MRST99 is $1.25\,\GeV^2$ and that's what we
set our value of $Q^2_f$ to be. 

\begin{figure}[t]
\epsfxsize=10.0cm
\begin{center}
\epsfbox{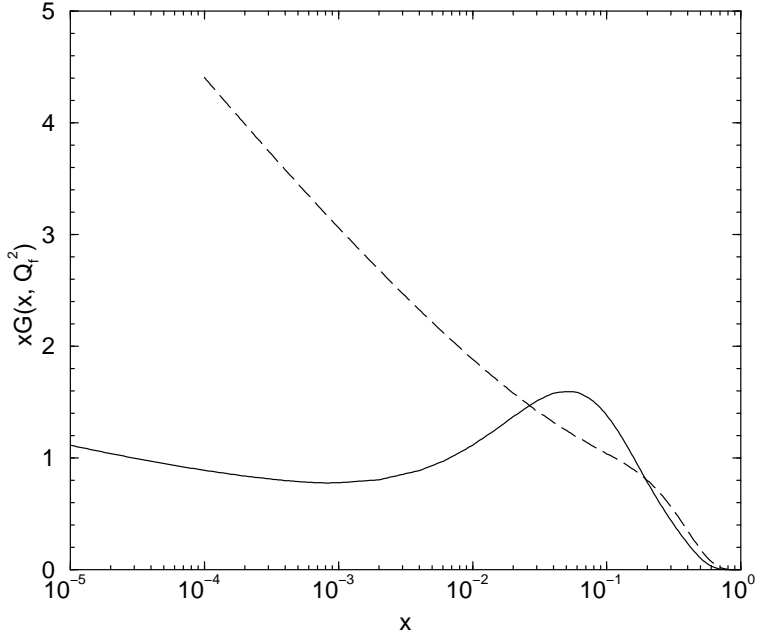}
\end{center}
\caption{
The gluon distribution function $xG(x,Q^2_f)$ from MRST99 (solid line) and
GRV98 (dashed line).  Both are calculated at $Q_f^2 = 1.25\,\GeV^2$.
}
\label{fig:xg}
\end{figure}
Using MRST99 the cross section integrated between $10^{-5} < x_L < 0.1$
is  
\be
\sigma = {1\over 2}(\sigma_{++} + \sigma_{+-}) = 0.21\,\hbox{mb}
\ee
while using GRV98 gives  
\be
\sigma = {1\over 2}(\sigma_{++} + \sigma_{+-}) = 1.0\,\hbox{mb}
\ee
in the same range. This difference in the cross-section 
is mainly due to the difference of the two gluon
distribution functions at small $x$.  For illustration, we plot 
$xG(x,Q^2_f)$ for both sets in Figure (\ref{fig:xg}).  

From the thermal model as well as exclusive reaction 
studies\cite{Castoldi:1997fn,Alde:1988kk}
one would expect the $\eta'$ cross-section to be
a few\,\% of the pion production cross-section, or 
$O(0.1\,\hbox{mb})$.  However, this does not necessarily imply that
MRST99 is better suited for our purpose.  To the authors' knowledge, 
there is no measurement of $\eta'$ inclusive production cross-section.  
By measuring the $\eta'$ inclusive production cross-section, one can 
better constraint the behavior of the gluon
distribution function at small $x$. 

The polarized gluon distributions from LSS2001 and GRSV2000 are
also very different.
Using LSS2001, we get 
\be
\Delta \sigma = (\sigma_{++} - \sigma_{+-})
= 0.97\,\mu\hbox{b} 
\ee
integrated in the $10^{-5} < x_L < 0.1$ range 
while using GRSV2000 gives 
\be
\Delta \sigma = (\sigma_{++} - \sigma_{+-})
= 0.17 \mu\hbox{b} 
\ee
in the same range. These results indicates that the asymmetry 
$A_{LL}$ can range from 
$1.7\times 10^{-4}$ to $4.6 \times 10^{-3}$ or can differ by
a factor of 30 or more depending on the choice of current gluon
distribution functions.  
Therefore measuring $\eta'$ cross-section and the asymmetry 
accurately can greatly help constrain different gluon distribution
functions currently in use.

\begin{figure}[htb]
\epsfxsize=10cm
\begin{center}
\epsfbox{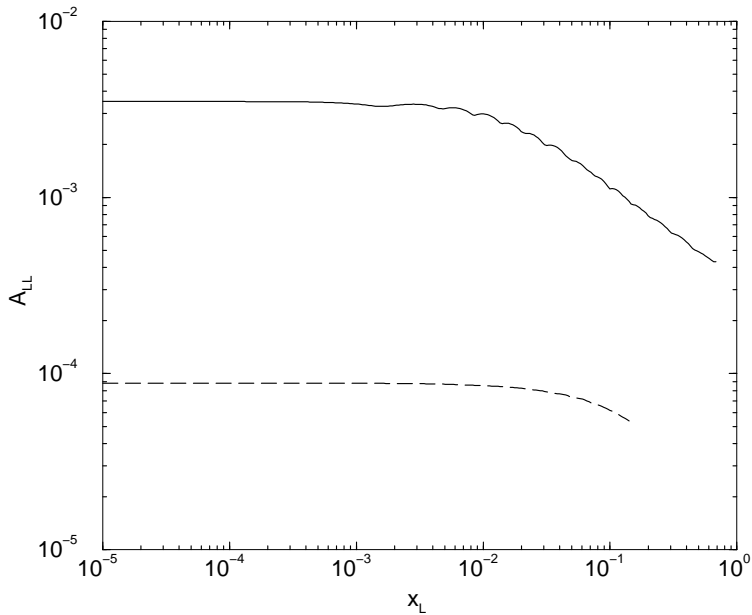}
\end{center}
\caption{
The asymmetry $A_{LL}$ as a function of $\eta'$ $x_L$ calculated at
$\sqrt{s}=250\,\GeV$ and $Q_f^2 = 1.25\,\GeV^2$. 
The upper curve is calculated using MRST99 and
LSS2001 parton distribution functions.  The lower curve is calculated
using GRV98 and GRSV2000 sets.  
}
\label{fig:ALLx}
\end{figure}

Figure (\ref{fig:ALLx}) shows $A_{LL}$ as a function of $\eta'$ meson 
momentum fraction $x_L$ at $\sqrt{s} = 250\,\GeV$.
Again we use the two sets of gluon distribution functions as described 
above.
The solid curve in Figure (\ref{fig:ALLx}) uses Set 1 (MRST99+LSS2001)
and 
the dashed curve uses Set 2 (GRV98+GRSV2000).
The difference between two sets is quite large.  The values of
$A_{LL}$ at small $x$ differ by a factor of almost 40. 

One notable feature of this figure is that for $x_L < 0.001$, $A_{LL}$
is constant.
To understand the plateau, we note that
\be
 x_\pm
 =
 {\sqrt{x_L^2 + 4M_{\eta'}^2/s} \pm x_L\over 2}
 \label{eq:xpm_sol}
\ee
If $x_L \ll M_{\eta'}/\sqrt{s}$, then 
$x_+ \approx x_- \approx M_{\eta'}/\sqrt{s}$ and
hence
\be
A_{LL} \approx 
\left(
{\Delta G(x=M_{\eta'}/\sqrt{s})\over G(x=M_{\eta'}/\sqrt{s})}
\right)^2
\label{eq:ALL_near_0}
\ee
will remain almost constant.  The value of the asymmetry $A_{LL}$ at 
$y = 0$ of course is a function of the collision energy.
The input parametrization used for the unpolarized gluon distribution
function for both LSS2001 and GRSV2000
is of the form
\be
x\Delta G(x) = C\, x^a\, (1-x)^b\, xG(x)
\label{eq:DeltaG}
\ee
If our $Q_f^2 = 1.25\,\GeV^2$ is not very much different from the input
$Q_0^2$ of the distribution functions, 
the asymmetry should be given by
\be
A_{LL}(x_L)
& \approx &
C^2\, x_+^a\, (1-x_+)^b\, x_-^a (1-x_-)^b
\non
& = &
C^2\left( M_{\eta'}^2\over s \right)^a (1-x_+)^b\,(1-x_-)^b
\ee
where we used $x_+ x_- = M_{\eta'}^2/s$.
Near $x_L = 0$, both $x_+$ and $x_-$ are small so that 
\be
A_{LL} \approx 
C^2 \left(M_{\eta'}^2\over s\right)^a
\label{eq:ALL_at_0}
\ee
Therefore measuring the asymmetry at $y=0$ for various energies can tell
us what the exponent $a$ should be {\em without knowing} the detailed
form of $xG(x)$. For LSS2001, this exponent turned out to
be about $0.6$ and for GRSV2000, it is about $0.8$.  
The reason that Set 2 curve deviates from a straight line may be
explained by the fact that the input $Q_0^2$ of set $2$
$(0.4\,\GeV^2)$ is not so close to our $Q_f^2 = 1.25\,\GeV^2$.
\begin{figure}[htb]
\epsfxsize=10cm
\begin{center}
\epsfbox{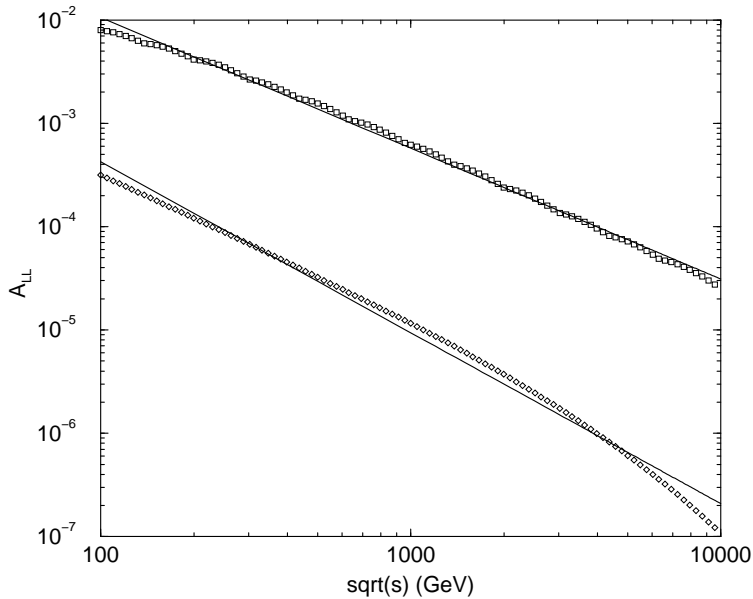}
\end{center}
\caption{The asymmetry $A_{LL}$ at the mid-rapidity calculated at $Q_f^2
= 1.25\,\GeV^2$.
MRST99 and LSS2001 are used for the upper curve and
GRV98 and GRSV2000 are used for the lower curve. 
The solid lines are linear fits.  For MRST99+LSS2001, the slope turned
out to be $-1.27$ corresponding to $a = 0.633$ in
Eq.(\protect\ref{eq:DeltaG}).  For GRV94+GRSV, the slope is
$-1.65$ corresponding to $a = 0.827$.
%
%
}
\label{fig:ALLs}
\end{figure}

Even though the detectors at RHIC can only go up to the maximum $x$ of
about 0.01, it is worth noting the behavior of $A_{LL}$ for larger
values of $x$ in Figure (\ref{fig:ALLx}).
When $x_L \gg M_{\eta'}/\sqrt{s}$,
$x_\pm$ can be approximated to be\footnote{
For RHIC at $\sqrt{s} = 250\,\GeV$, 
$M_{\eta'}/\sqrt{s} = 0.0038$.  For LHC at $\sqrt{s} = 5500\,\GeV$,
$M_{\eta'}/\sqrt{s} = 0.00017$.  
}
\be
 x_+ \approx x_L + {M_{\eta'}^2\over s x_L}
 \ \ \ \hbox{and}\ \ \  
 x_- \approx {M_{\eta'}^2\over sx_L} 
\ee
and the asymmetry becomes
\be
A_{LL}(x_L)
& \approx &
C^2\left( M_{\eta'}^2\over s \right)^a 
\left(1-x_L-{M_{\eta'}^2\over s x_L}\right)^b\,
\left(1-{M_{\eta'}^2\over sx_L}\right)^b
\ee
In LSS2001, the exponent $b$ is argued to be 0.  In other
parametrizations, $b$ is substantially different from 0.
For instance in GRSV2000, $b$ ranges from 4 to 7.
Therefore even at moderately large $x_L$, the behavior of $A_{LL}$ as a
function of $x_L$ depends strongly on the exponent $b$. 
In this way, very forward $\eta'$'s can distinguish between different 
parameterizations of the unpolarized gluon distribution function.

\section{Conclusion}

In this paper, we showed the dependence of $A_{LL}$ 
on $\eta'$ $x_L$ as well as the ($pp$) center of mass energy of the
collison using (\ref{eq:Aresult}).  Measuring $A_{LL}$ in this way
directly probes the gluon distribution function in the proton.
Had we known the unpolarized gluon distribution function well, 
$A_{LL}$ could be used to map out the polarized gluon distribution.
As it happens, currently the gluon distribution is not very well
constrained by the past experiments in the small $x$ region.
Measuring $\eta'$ at high energy can greatly constrain 
the gluon distribution in this region.

There are several caveats to our results. First and perhaps 
the most important one theoretically, is that $M_{\eta'} \sim 1 \GeV$ 
is not very large
and therefore higher order (in $\alpha_s$) corrections can be
quite sizable. This is somewhat ``cured'' by introducing a $K$
factor which is supposed to take higher order corrections into effect.
In our case, since we are calculating the ratio of cross sections,
these effects should largely cancel out. However, without a real calculation
of higher order effects, there is no way to know for sure. Another
complication arises due to effects such as initial state radiation
of gluons, etc. which would lead to the produced particle having
a transverse momentum. These effects can be resummed for few
limited processes but are mostly included phenomenologically by
introducing an intrinsic $k_t$ for the incoming partons \cite{apa}.
We intend to do this in the near future and investigate the
dependence of $A_{LL}$ on the transverse momentum of $\eta'$.
However, if we consider $p_t$ smaller than the intrinsic $k_t$ scale,
our analysis should still apply.

Another potential problem is that we have neglected
possible contribution of annihilation of quarks and anti quarks
into $\eta'$s. We believe we can avoid this problem here for two
reasons.  First,
the values of $x$ considered are quite small where gluons are
the dominant partons. Therefore contribution of sea quarks to this
process should be of order of a few percent which we can safely neglect. 
Furthermore, if the $q\bar q \eta'$ vertex is of the derivative
type\cite{Muta:1999ue},
then in the on-shell limit the quark anti-quark annihilation 
matrix element is $O(m_q/M_{\eta'})$ smaller than the gluon fusion
matrix element.  Hence, in the small $x$ range the gluon fusion process
should dominate over quark anti-quark annihilation process.

In summary, we have investigated the production of $\eta'$ in 
polarized $pp$ collisions and estimated the double spin asymmetry
variable $A_{LL}$ using the available polarized gluon distribution
functions. Alternatively, measuring $A_{LL}$
at polarized proton-proton collisions at RHIC will enable us to
directly extract the polarized gluon distribution function at
$Q^2\sim 1$ and different values of $x$.

\section*{Acknowledgement}

We would like to thank G.~Bunce, K.~Itakura and W.~Vogelsang for 
useful discussions.  We also thank W.~Vogelsang for providing us with 
the GRSV2000 FORTRAN code.
J.J-M.~is supported in part by a PDF from BSA and by U.S. Department 
of Energy under Contract No. DE-AC02-98CH10886.
S.J.~was in part supported by the National Science Foundation,
Grant No. PHY-00-70818, by the Natural Sciences and 
Engineering Council of Canada and by le Fonds pour la Formation 
de Chercheurs et l'Aide \`a la Recherche du Qu\`ebec.

\end{document}